\documentclass[12pt,a4paper]{article}

\usepackage{amsmath,amssymb,amsfonts,mathrsfs,amsthm,mathtools}
\usepackage{physics}
\usepackage{tensor}
\usepackage{graphicx}
\usepackage{caption}
\usepackage{subcaption}
\usepackage{geometry}
\usepackage{setspace}
\usepackage{newtxtext,newtxmath}
\usepackage[hidelinks]{hyperref}
\usepackage[backend=biber,style=numeric,sorting=none]{biblatex}
\usepackage{float}
\title{\textbf{Stochastic Quantum Mechanics Trajectories Near Schwarzschild Horizon Black Holes}}

\author{
Juan S. Jerez- Rodríguez\textsuperscript{1,}\thanks{\href{mailto:juan.jerez@cinvestav.mx}{juan.jerez@cinvestav.mx}},
Eric S. Escobar-Aguilar\textsuperscript{2,}\thanks{\href{mailto:e.escobar@xanum.uam.mx}{e.escobar@xanum.uam.mx}},
and Tonatiuh Matos\textsuperscript{1,}\thanks{\href{mailto:tonatiuh.matos@cinvestav.mx}{tonatiuh.matos@cinvestav.mx}}\\[6pt]
\textsuperscript{1}\textit{Departamento de Física, Centro de Investigación y de Estudios Avanzados,}\\
\textit{Av. Instituto Politécnico Nacional 2508, 07360 CDMX, Ciudad de México, México}\\[3pt]
\textsuperscript{2}\textit{Departamento de Física, Universidad Autónoma Metropolitana Iztapalapa,}\\
\textit{San Rafael Atlixco 186, 09340 CDMX, Ciudad de México, México}
}

\date{(Dated: \today)}

\begin{document}

\maketitle

\begin{abstract}
This work explores the possibility of applying stochastic quantum mechanics to curved spacetimes, with an emphasis on the Schwarzschild black hole. After reviewing the fundamental concepts of this approach, the quantum stochastic equations are extended to curved spacetime using a fully covariant treatment. Subsequently, the Klein-Gordon equation is solved for scalar perturbations, and the resulting stochastic trajectories are analyzed by varying parameters such as angular momentum, particle frequency, and computational integration time. In conclusion, we find that the trajectories are influenced by gravitational fluctuations in spacetime and that, depending on the variation of the fundamental parameters, different types of stochastic trajectories are obtained.
\end{abstract}

\section{Introduction}

It is a well-established fact that we live immersed in a Background of Gravitational Waves (GWB). Strictly speaking, all motion in the universe generates gravitational waves: from the Big Bang, inflation, the production of primordial black holes, through the formation and collision of black holes, to the collective motion of galaxies (see \cite{Agazie_2023,Reardon_2023,Xu_2023,refId0}). These fluctuations in spacetime are extremely small; however, if the energy of massless excitations satisfies Planck's relation $E=h \nu=hc/\lambda$, where $h$ is Planck's constant and $\lambda$ is the wavelength of the particle, even very small fluctuations can carry large amounts of energy when their wavelength is sufficiently short. Therefore, it is worthwhile to consider spacetime fluctuations as a relevant component in particle dynamics. This idea is similar to that of a lake where boats don't perceive the small fluctuations, but ants or fleas do perceive the movement of the waves.

There are several ways to model the gravitational wave background. One possibility is to treat the GWB as a spacetime with stochastic fluctuations and incorporate these fluctuations directly into the gravitational description, as occurs in certain recent approaches based on master equations for quantum dynamics in a classical gravitational background \cite{PhysRevX.13.041040}. By contrast, following the approximation introduced by Escobar, Matos, and Aquino (see \cite{escobar2025fundamental})—which forms the basis of this work—we will assume that the GWB produces only a minimal alteration of the geometry, allowing us to ignore stochasticity in Einstein's field equations, but incorporating the randomness to the particle's dynamics. From this perspective, it is shown that when considering the geodesic of a particle with an added stochastic term in an arbitrary curved spacetime, the dynamical equation that the particle satisfies coincides with the Klein–Gordon equation for the complex function $\Phi=\sqrt{n}\exp(\Theta)$. It was further shown that the phase $\Theta$ determines the geometry for the particle in spacetime, while $n$ adds stochastic behavior through the velocity. Since the non-relativistic limit of the Klein-Gordon equation is the Schrödinger equation, this framework implies that quantum mechanics (QM) can be determined from the stochastic nature of spacetime fluctuations, rather than from the uncertainty principle. In this way we provide an alternative interpretation of quantum mechanics.

Black holes provide an ideal environment to test these ideas. As extreme solutions of Einstein’s field equations, they generate strong gravitational fields and highly curved regions where even small stochastic effects may be amplified. The Schwarzschild solution, discovered in 1916, remains the simplest and most studied black hole geometry. Since then, major advances—from Chandrasekhar’s theory of gravitational collapse to the Kerr solution and Penrose’s singularity theorem—have shaped our current understanding of black holes. Recent observational milestones, including the discovery of a supermassive black hole at the center of the Milky Way \cite{PhysRevLett.124.081101,2008ApJ...689.1044G,RevModPhys.82.3121} and the Event Horizon Telescope images of M87* and Sgr A* \cite{Akiyama_2019_short, EHT_SgrA_2022},  provide confirmation of these objects and strengthen the motivation to explore quantum behavior in their vicinity.

This work explores stochastic quantum mechanics in the context of curved spacetime, with a particular focus on the Schwarzschild metric. From exact solutions to the Klein-Gordon equation, we determine both the geodesic and stochastic velocities of particles, allowing us to calculate and analyze their trajectories in the vicinity of a black hole and represent them in different frames of reference. The article is organized as follows:

Section \ref{ch2} presents the covariantization of stochastic quantum equations following the methodology presented in \cite{chavanis2016derivationgeneralizedschrodingerequation, axioms13090606}, but since the treatment is in the relativistic regime we aim for a relativistic generalization based primarily on the results of Escobar, Matos, and Aquino \cite{escobar2025fundamental}. In Section \ref{ch3}, we study the solution of the Klein–Gordon equation for massless scalar fields in Schwarzschild spacetime. The application of these solutions within stochastic quantum mechanics in curved spacetimes is developed in Section \ref{ch4}, where we analyze the stochastic trajectories of highly energetic photons by varying specific parameters such as angular momentum $l$, particle's frequency $\omega_{0}$, and computational integration time. Finally, in Section \ref{ch5}, we discuss and draw conclusions about the results obtained.

%%%%%%%%%%%%%%%%%%%%%%%%%%%%%%%%%%%%%%%%%%%%%%%%%%%%%%%%%%%%%%%%%%%%%%%%%%%%%%%%%%%%%%%%%%%%%%%%%%%%%%%%%%%%%%%%%%%%%%%%%%%%%%%%%%%%%%%%%%%%%%%%%%%%%%%%%%%%%%%%%%%%%%%%%%%%%%%%%%%%%%%%%%%%%%%%%%%%%%%%%%%%%%%%%%%%%%%%%%%%%%%%%%%%%%%%%%%%%%%%%%%%%%%%%
\section{Stochastic Quantum Mechanics in curved spacetime}
\label{ch2}
In this section we set the background necessary to the stochastic treatment for the quantum particles. 
The main objective is to show the equivalence between stochastic and quantum mechanics. In particular, both formulations gather in the hydrodynamic representation, that includes complex velocity fields, and Hamilton–Jacobi–type equations which have been discussed in \cite{chavanis2016derivationgeneralizedschrodingerequation,axioms13090606}, we follow this procedures as part of the theoretical foundation of this work. The stochastic formalism begin by considering that the particles are surrounded by an stochastic gravitational Background, that may be given by the GWB, as this background is small then we consider that only sufficiently small particles are affected. The particles dynamics changes as it lives in a stochastic background, instead of describing geodesics, the particles trajectories are stochastic trajectories. 

To model this trajectories we consider the formalism described in \cite{escobar2025fundamental}, where the relation between quantum mechanics and stochastic processes in curved spacetime is proved. The formalism consider the stochastic differential equation given by the definition of the 4-velocity $\mathcal{U}^\mu$, and we add a stochastic term
%%%%%%%%%%%%%%%%%%%%%%%%%%%%%%%%%%%%%%%%%%%%%%%%%%%
\begin{equation}
   \frac{d x^\mu}{d\tau}=\mathcal{U}^\mu + \sqrt{2\sigma}\, \xi^{\mu}(\tau), \label{eq:4velocity}
\end{equation} 
%%%%%%%%%%%%%%%%%%%%%%%%%%%%%%%%%%%%%%%%%%%%%%%%%%
where $x^\mu(\tau)$ is the world-line of the particle and $\tau$ is the proper time, the second term on the right side represents the stochastic contribution, being $\sigma$ the intensity of the noise and $\xi^\mu(\tau)$  a random variable with statistical properties of a Gaussian White Noise (GWN) with the statistical properties given by: zero mean value $\langle \xi^{\mu}(\tau)\rangle=0$ and correlation function $\langle\xi^\mu(\tau_1)\xi^\nu(\tau_2)\rangle= \delta^{\mu\nu} \delta(\tau_2-\tau_1)$. The stochastic term causes particles to deviate from classical deterministic paths, resulting in stochastic motion.

In stochastic processes time-reversal can be treated as two different stochastic processes one representing forward time evolution $\hat{W}_{+}^{\mu}(\tau)$ and the other representing backward time evolution $\hat{W}_{-}^{\mu}(\tau)$. The mathematical formulation allows predictions about both future and past states based on the current state. With this is suitable to write Eq. (\ref{eq:4velocity}) in differential notation as
\begin{align}
dx_{+}^{\mu}=\mathcal{U}_{+}^{\mu} d  \tau + \sqrt{2 \sigma} d \hat{W}_{+}^{\mu}(\tau),\label{eq:Forward SDE}\\
dx_{-}^{\mu}=\mathcal{U}_{-}^{\mu} d  \tau + \sqrt{2 \sigma} d \hat{W}_{-}^{\mu}(\tau).\label{eq:Backward SDE}
\end{align}
%%%%%%%%%%%%%%%%%%%%%%%%%%%%%%%%%%%%%%%%%%%%%%%%
Where $d\hat{W}^\mu_+(\tau)$ and $d\hat{W}^\mu_-(\tau)$ satisfy the properties of Gaussian white noise described before, the zero mean value
\begin{equation} 
    \langle d\hat{W}^\mu_+(\tau) \rangle=\langle d\hat{W}^\mu_-(\tau) \rangle=0, \label{mean zero}
\end{equation} 
and correlation function 
\begin{equation}
    \langle d \hat{W}^{\mu}_{\pm}(\tau) d \hat{W}^{\nu}_{\pm}(\tau')\rangle= \pm d\tau d\tau'  \delta(\tau- \tau') \delta^{\mu \nu}. \label{correlation func}
\end{equation}
%%%%%%%%%%%%%%%%%%%%%%%%%%%%%
The $\delta(\tau-\tau')$ is called Markovian property, meaning that the processes at different times are independent, however, the variance when $\tau=\tau'$ is $\langle (d\hat W^\mu(\tau))d\hat{W}^\nu(\tau)\rangle= \delta^{\mu\nu} d\tau$. In this case, it is said that the Wiener process is a Markovian and Gaussian stochastic process, the same as the dynamic variable $x^\mu(\tau).$  The fact that we have two stochastic differential equations lead to different behavior in the dynamics, such dynamics should be a manifestation of the different forces and spacetime itself. It is clear that we should be able to find the invariant generalization of the equation of motion
\begin{equation}
    a^\mu = f^\mu/\mu_0,
\end{equation}where $a^\mu$ is the acceleration, $f^\mu$ are the forces that act on the particle, and the mass parameter is $\mu_0.$ Since we want to model quantum particles, we can consider also massless particles, in such case we consider the mass parameter proportional to the frequency of the particle $\omega_{0}$.

As we will show latter, the stochastic trajectories are continuous but non-differentiable, however since we are modeling this trajectories with stochastic processes we can use the Itô Calculus. The main result, and pillar of the stochastic calculus is the Ito's Lemma, where $\langle (d\hat W^\mu(\tau))d\hat{W}^\nu(\tau)\rangle= \delta^{\mu\nu} d\tau.$ This fact lead to a general differential operator given by

\begin{equation}\label{D-decomp}
\hat{D} = \hat{D}_c - i\,\hat{D}_s,
\end{equation}
where adopt the current derivative is given by
\begin{equation}\label{Dc-Ds}
\hat{D}_c = \pi^\alpha \nabla_\alpha,
\end{equation}and the stochastic derivative is 
\begin{equation}
    \hat{D}_s = u^\alpha \nabla_\alpha + \sigma\,\nabla_\alpha\nabla^\alpha,
\end{equation}
where \(\pi^\alpha\) and \(u^\alpha\) are given velocity fields obtained by applying the differential operator on $x^\mu,$ considering that as we are dealing with quantum particles the vicinity spacetime regions to the particles are locally flat so $\nabla_\nu x^\mu=\delta^\mu_\nu,$ then
\begin{equation}
     \hat{D}_c x^\mu = \pi^\mu,
 \end{equation} and 
 \begin{equation}
     \hat{D}_s x^\mu = u^\mu.
 \end{equation} So the generalized velocity is given by
 \begin{equation}
     \omega^\mu = \hat{D} x^\mu = \pi^\mu - i  u^\mu.
 \end{equation}
 For the acceleration we have the differential operator $\hat{D}$ acting on \(\omega^\mu\) defining the generalized acceleration as 
 \begin{equation}
     a^\mu=\hat{D} \omega^\mu = \omega^\alpha\nabla_\alpha\omega^\mu - i\,\sigma\,\nabla_\alpha\nabla^\alpha\omega^\mu.
 \end{equation}

Suppose the acceleration is equated to a source-like force in the right hand side involving a generalized force $f^\mu= f_+^\mu - if_-^\mu,$ as stated in \cite{escobar2025fundamental} we have that $f_+^\mu=-\mu_0\nabla^{\mu}\mathcal{A}$ and  $f_-^\mu = 0$ in absence of electromagnetic force. Then we have 
\begin{equation}
\hat{D}\,\omega^\mu =-\nabla^{\mu}\mathcal{A} \label{generalDynamicequation},
\end{equation}where $\mathcal{A}$ is a potential that parametrize the self-interacting potential $V$ in the classical Klein-Gordon equation. 

Following \cite{chavanis2016derivationgeneralizedschrodingerequation,axioms13090606}, we determine the generalized form for the velocity $\omega^\mu$ resembling the Lagrangian mechanics which states that given a complex scalar action \(\mathcal{S}\), we can define the complex momentum as
\begin{equation}\label{momenta}
p^\mu := \nabla^\mu \mathcal{S},
\end{equation}
Then we adopt the ansatz for the velocity as
\begin{equation}\label{omega-S}
\omega^\mu = \frac{\nabla^\mu \mathcal{S}}{\mu_0}.
\end{equation}
Substituting \eqref{omega-S} into \eqref{generalDynamicequation} yields

\begin{equation}\label{pre-contraction}
\frac{1}{\mu_0}\bigl(\nabla^\alpha \mathcal{S}\bigr)\bigl(\nabla_\alpha\nabla^\mu \mathcal{S}\bigr)
- i\,\sigma\,\nabla_\alpha\nabla^\alpha\!\bigl(\nabla^\mu \mathcal{S}\bigr)
= -\mu_0\nabla^{\mu}\mathcal{A}.
\end{equation}

To reduce the vector equation to an (effective) scalar equation, we integrate (under standard manipulations and appropriate boundary/regularity assumptions) into the scalar equation
\begin{equation}\label{scalar-master}
\frac{1}{2\mu_0}\nabla^\alpha \mathcal{S}\,\nabla_\alpha \mathcal{S} - i\,\sigma\,\nabla_\alpha\nabla^\alpha \mathcal{S} = -\mu_0 \mathcal{A}.
\end{equation}
The procedure followed is aligned with the hydrodynamic form of the quantum equations, (see \cite{chavanis2016derivationgeneralizedschrodingerequation, axioms13090606, Madelung1927QuantentheorieIH, Matos_2019}). With this version of the Klein-Gordon equation we can identify the former equation as a Hamilton-Jacobi equation, or a Bernoulli equation. 
To explicitly obtain the Klein-Gordon equation we introduce the Cole-Hopf transformation, as in the procedure done in \cite{chavanis2016derivationgeneralizedschrodingerequation}, where the wave function appears as
\begin{equation}\label{colehopf}
\mathcal{S} = -2\,i\,\sigma \mu_0 \ln\Phi,
\end{equation}
where \(\Phi\) is a scalar field (assumed nonzero). We compute the necessary derivatives of \(\mathcal{S}\):

\begin{equation}\label{gradS}
\nabla_\alpha \mathcal{S} = -2 i \mu_0\,\sigma\frac{\nabla_\alpha\Phi}{\Phi}.
\end{equation}

\begin{equation}\label{gradS-squared}
\nabla^\alpha \mathcal{S}\,\nabla_\alpha \mathcal{S}
= (-2 i \mu_0\sigma)^2\,\frac{\nabla^\alpha\Phi\,\nabla_\alpha\Phi}{\Phi^2}
= -4\mu_0^2\sigma^2\,\frac{\nabla^\alpha\Phi\,\nabla_\alpha\Phi}{\Phi^2}.
\end{equation}
Using the identity
\begin{equation}\label{lap-log}
\Box\ln\Phi = \frac{\Phi\,\Box\Phi - \nabla^\alpha\Phi\,\nabla_\alpha\Phi}{\Phi^2},
\end{equation}where $\Box = \nabla_\alpha\nabla^\alpha$
we find that eq. \eqref{scalar-master} results in

\begin{equation}\label{combined}
\nabla^\alpha\,\nabla_\alpha\Phi
= \frac{-\mathcal{A}\Phi}{2\mu_0\sigma^2}.
\end{equation}

We introduce the physically motivated identification, as this work just for quantum particles \cite{Nelson1967DynamicalTO,escobar2025fundamental,chavanis2016derivationgeneralizedschrodingerequation}, then 
\begin{equation}\label{sigma-hbar}
\sigma = \frac{\hbar}{2\mu_0}.
\end{equation}
With this choice, we finally get
\begin{equation}\label{nonlinear-phi-hbar}
\Box\Phi = -\frac{\hbar\mathcal{A}\Phi}{2\mu_0}
\end{equation}
Now, if we consider free particles $\mathcal{A}=0,$ then we have the free particle, massless Klein-Gordon equation

\begin{equation}\label{K-G eq}
\Box\Phi = 0.
\end{equation} 

The form of $\mathcal{S}$ is in general a complex action, an we proved previously that the case in which the action is $\mathcal{S}=S - \omega_0 \hbar x^0 - \frac{i\hbar}{2}\ln n$, we recover the Madelung transformation where $n$ is the particle's density, $x^0$ is the evolution parameter in the metric, $\omega_0$ is either the mass or the frequency of the particle and $S$ encodes the geometry in the phase. The Madelung transformation is then written as
\begin{equation}
\Phi(t, \mathbf{x}) = \sqrt{n} e^{\frac{i}{\hbar}(S - \frac{\omega_0 \hbar}{c} x^0)}. \label{eq:madelungUnits}
\end{equation}With this choice we can easily find the generalized velocity given by
\begin{equation}
    \omega^\mu= \frac{\nabla^\mu\mathcal{S}}{\mu_0}= \nabla^\mu\frac{1}{\mu_0}\left(\nabla^\mu S - \omega_0 \hbar \nabla^\mu x^0 -  \frac{i\hbar}{2}\nabla^\mu(\ln n) \right). \label{total velocity}
\end{equation}
Here we can identify two velocities, one depending only on the phase, normally associated with the current velocity
\begin{equation} \label{Pi sin e}
    \pi^\mu= \frac{\nabla^\mu S}{\mu_0} - \frac{\omega_0 \hbar}{\mu_0} \nabla^\mu x^0,
\end{equation}
the second velocity, depending on the amplitude, is identified with a stochastic velocity  defined as
\begin{equation}\label{u relativista}
    u^\mu=\frac{\hbar}{2\mu_0}\nabla^\mu(\ln n).
\end{equation}With these two velocities we can now draw the stochastic trajectories in a certain spacetime either by using the forward or backward stochastic differential equations
\begin{align}
dx_{+}^{\mu}=(\pi^\mu + u^\mu) d  \tau + \sqrt{2 \sigma} d \hat{W}_{+}^{\mu}(\tau),\label{eq:Forward SDE complete}\\
dx_{-}^{\mu}=(\pi^\mu - u^\mu) d  \tau + \sqrt{2 \sigma} d \hat{W}_{-}^{\mu}(\tau).\label{eq:Backward SDE complete}
\end{align}
Then we have the mechanics that lead to the stochastic trajectories, however the dynamics lead to velocities that are in terms of $S$ and $n,$ and those terms are related to the solution of the wave equation given by (\ref{K-G eq}). This is an important feature of the formalism, since the stochastic differential equations lead only to the kinematic form of the equations and the dynamics must consider the background, and its geometry, and solving the Klein-Gordon equation we obtain just the deterministic part of the stochastic differential equation.
To resume this section we proved that the Klein-Gordon equation, in this case for the free-particle can be obtained from an stochastic formalism, and with this close relationship between stochastic processes and quantum mechanics we are able to obtain quantum trajectories in a given spacetime configuration.

%%%%%%%%%%%%%%%%%%%%%%%%%%%%%%%%%%%%%%%%%%%%%%%%%%%%%%%%%%%%%%%%%%%%%%%%%%%%%%%%%%%%%%%%%%%%%%%%%%%%%%%%%%%%%%%%%%%%%%%%%%%%%%%%%%%%%%%%%%%%%%%%%%%%%%%%%%%%%%%%%%%%%%%%%%%%%%%%%%%%%%%%%%%%%%%%%%%%%%%%%%%%%%%%%%%%%%%%%%%%%%%%%%%%%%%%%%%%%%%%%%%%%%%%%

\section{Klein-Gordon Equation with a Massless Scalar Field in a Schwarzschild Background}
\label{ch3}
The main objective in this section is to observe how scalar fields behave in a Schwarzschild black hole background. Specifically, we focus on a scalar field  $\Phi$ that is massless, does not interact with itself, and is minimally coupled to the background geometry. The dynamics of this field are described by the Klein-Gordon equation
\begin{equation}
    \Box \Phi(ct,\textbf{r}) \equiv \nabla^{\mu} \nabla_{\mu}  \Phi(ct,\textbf{r})  = \sqrt{-g} \dfrac{\partial}{\partial x^{\mu}} \left[ 
 \sqrt{-g} g^{\mu \nu} \dfrac{\partial}{\partial x^{\nu}}\right] \Phi(ct,\textbf{r})=0
\label{eq:2-1}
\end{equation}
where $g$ is the determinant of the Schwarzschild metric
\begin{equation}
    ds^{2}= -\left(1-\dfrac{r_{s}}{r}\right) d(ct)^2+ \left( 1-\dfrac{r_{s}}{r}\right)^{-1}dr^2+ r^{2} d\theta^{2}+ r^{2} \mbox{sin}^{2}\theta d\phi^{2},
\end{equation}
where the Schwarzschild radius is defined as $r_{s}= 2MG/c^{2}$, with $M$ being the mass of the black hole, $G$ the gravitational constant and $c$ the velocity of light. The metric signature adopted here is $(-,+,+,+)$.

Let us assume that the energy associated with the scalar field configuration is much smaller than the mass of the black hole (test-field limit), so that its gravitational backreaction can be safely neglected. In this regime, the entire geometric description is determined solely by the black hole metric. Expanding Eq. (\ref{eq:2-1}), we find
\begin{equation}
-\left(1-\dfrac{r_{s}}{r}\right) \dfrac{\partial^{2} \Phi}{\partial t^{2}} \dfrac{1}{c^{2}} + \dfrac{1}{r^{2}} \left[\dfrac{\partial}{\partial r}\left[r^{2}\left(1-\dfrac{r_{s}}{r}\right) \dfrac{\partial \Phi}{\partial r}\right]\right]+ \dfrac{1}{r^{2} \sin\theta} \left[\dfrac{\partial }{\partial \theta}\left(\sin\theta \dfrac{\partial \Phi}{\partial \theta} \right) \right]+ \dfrac{1}{r^{2} \sin^{2} \theta} \dfrac{\partial^{2} \Phi}{\partial \phi^{2}}=0.
\label{eq:2-2}
\end{equation}
Given that we have a spherically symmetric metric, it is natural to propose a separable solution of Eq. (\ref{eq:2-2}) in the form
\begin{equation}
\Phi(ct,\textbf{r})= \dfrac{\Psi_{l}(ct,r)}{r} Y_{l}^{m}(\theta,\phi),
\label{eq:2-3}
\end{equation}
being $Y_{l}^{m}$ the spherical harmonics. Substituting Eq. (\ref{eq:2-3}) into Eq. (\ref{eq:2-2}), we obtain
\begin{equation}
\dfrac{1}{c^{2}} \left( \dfrac{\partial^{2}\Psi}{\partial t^{2}}\right)- \dfrac{r_{s}}{r^{2}}\left(1-\dfrac{r_{s}}{r}\right)\left( \dfrac{\partial \Psi}{\partial r}\right)- \left(1-\dfrac{r_{s}}{r}\right)^{2} \left(\dfrac{\partial^{2} \Psi}{\partial r^{2}} \right)+ \overbrace{\left(1-\dfrac{r_{s}}{r}\right)\left[
\dfrac{r_{s}}{r^{3}}+ \dfrac{l(l+1)}{r^{2}}\right]}^{V_{l}(r)}\Psi=0,
\label{eq:2-4}
\end{equation}
where the potential $V_{l}(r)$ is sometimes referred to as the "effective potential". A detailed numerical analysis of the radial equation in Eq. (\ref{eq:2-4}) using tortoise coordinates can be found in \cite{Barranco:2011eyw}. The solution to this equation describes wave propagation that can be analyzed in terms of both scattering and quasi-normal modes, phenomena influenced by the effective potential acting as a potential barrier (see \cite{philipp2015analyticsolutionswaveequations}). An exact analytical solution in terms of the confluent Heun function was obtained in  \cite{philipp2015analyticsolutionswaveequations}, and it is precisely this solution—and its associated analysis—that we employ in this work.

To further investigate the behavior of the solutions near the event horizon, it is instructive to examine the coordinate system itself. The original Schwarzschild coordinate system $(t,r)$ encounters a problem as one approaches the event horizon. As shown in the equation (\ref{eq:2-4}), both $r=0$ and $r=r_{s}$ — corresponding respectively to the black hole’s center and its event horizon — appear as singularities of the differential equation. However, the singularity at $r=r_{s}$ is merely a coordinate (or removable) singularity, meaning that the spacetime geometry remains regular there, allowing us to describe what happens when crossing the event horizon.

With this in mind, we now apply a general coordinate transformation to equation (\ref{eq:2-4}), where $(ct,r)\rightarrow (cv,u)$ and defining
\begin{equation}
    cv=f(ct,r)\mbox{, }u=g(ct,r).
\end{equation}
Both angular variables, $\theta$ and $\phi$, remain invariant, ensuring that the coordinates are consistent with spherical symmetry. In equation (\ref{eq:2-4}), the partial derivatives with respect to $r$ (for example) imply that $ct$ remains fixed (and vice versa). For our new coordinate system, we will use the respective partial derivatives of $u$ with $cv$ fixed (and vice versa). Substituting the transformed derivatives into equation (\ref{eq:2-4}), we obtain a differential equation in the coordinate system $(u,cv)$ of the form 
\begin{equation} \label{eq:2-5}
\begin{split}
&\left[\left(1- \dfrac{r_{s}}{r} \right)^{2} f'^{2}-\dot{f}^{2}\right] \dfrac{\partial^{2} \Psi(u,cv)}{\partial (cv)^{2}}\\
+&\left[\left(1- \dfrac{r_{s}}{r} \right)^{2} g'^{2}-\dot{g}^{2}\right] \dfrac{\partial^{2} \Psi(u,cv)}{\partial u^{2}} \\
+&\left[2\left(1- \dfrac{r_{s}}{r} \right)^{2} f'g'-\dot{f} \dot{g}\right] \dfrac{\partial^{2} \Psi(u,cv)}{\partial u \partial (cv)}\\
+& \left[\left(1- \dfrac{r_{s}}{r} \right)^{2} f''+ \dfrac{r_{s}}{r^{2}}\left(1- \dfrac{r_{s}}{r} \right) f'- \ddot{f} \right] \dfrac{\partial \Psi(u,cv)}{\partial (cv)}\\
+& \left[\left(1- \dfrac{r_{s}}{r} \right)^{2} g''+ \dfrac{r_{s}}{r^{2}}\left(1- \dfrac{r_{s}}{r} \right) g'- \ddot{g} \right] \dfrac{\partial \Psi(u,cv)}{\partial u}\\
-& V_{l}(r) \Psi(u,cv)=0,
\end{split}
\end{equation}
where $r$ is understood as a function of the null coordinates, $r=r(u,cv)$. At this stage, the functions $f(ct,r)$ and $g(ct,r)$ remain completely arbitrary. Therefore, equation (\ref{eq:2-5}) represents the general form of the transformed radial equation under any spherically symmetric coordinate transformation. We introduce the ingoing Eddington-Finkelstein (E-F) coordinates by specifying the two functions $f(r,t)$ and $g(r,t)$. The metric in terms of the coordinates $v$ looks like
\begin{equation}
    ds^{2}=- \left(1-\dfrac{r_{s}}{r} \right)d(cv)^{2}+ \left(d(cv)dr+drd(cv)\right)+r^{2}d\theta^{2}+r^{2}\mbox{sin}\theta d\phi^{2},
\label{eq2:5}
\end{equation}
and the corresponding transformation functions are
\begin{align}
    f=cv= ct +r + r_{s} \mbox{ln} \left| \dfrac{r}{r_{s}}-1\right| \label{eq-2-20},\\ 
    g=u\equiv r.
\label{eq-2-21}
\end{align}
Substituting equations (\ref{eq-2-20}) and (\ref{eq-2-21}) in the equation (\ref{eq:2-5}), we obtain
\begin{equation}
    \left[\left( 1-\dfrac{r_{s}}{r}\right)^{2} \dfrac{\partial^{2}}{\partial r^{2}} + 2 \left( 1- \dfrac{r_{s}}{r}\right) \dfrac{\partial^{2}}{\partial r \partial (cv)}+ \dfrac{r_{s}}{r^{2}}\left( 1-\dfrac{r_{s}}{r}\right) \dfrac{\partial }{\partial r}-V_{l}(r)\right]\Psi(cv,r)=0.
\label{eq:2-6}
\end{equation}
This equation separates under the ansatz
\begin{equation}
    \Psi(cv,r)= e^{-i \omega cv} R_{\omega l}(r),
\label{eq:2-7}
\end{equation}
and substituting (\ref{eq:2-7}) into (\ref{eq:2-6}) yields the radial equation 
\begin{equation}
    \left[\dfrac{d^{2}}{d r^{2}}+ \dfrac{r_{s}-2i \omega r^{2}}{r(r-r_{s})} \dfrac{d}{d r}- \left( \dfrac{l(l+1)+ \dfrac{r_{s}}{r}}{r(r-r_{s})}\right)\right]R_{\omega l}(r)=0.
\label{eq:2-8}
\end{equation}

This radial differential equation has regular singular points at $r=0$ (the curvature singularity) and $r=r_s$ (the event horizon), and an irregular singularity at $r=\infty$, as can be seen by setting $r=1/w$ and examining the limit $w\to 0$. With these properties, equation (\ref{eq:2-8}) belongs to the class of confluent Heun equations (see Appendix (\ref{chap: Appendix A})). To bring it into the standard confluent Heun form, we perform the following transformation:
\begin{equation}
    R_{\omega l}(r)= r y_{\omega l}(r).
\label{eq:2-8.1}
\end{equation}
For the rest of this discussion, we will adopt natural units and express all distances in terms of the Schwarzschild radius, $r_s = 2M$. With this choice, the event horizon is placed at $r = 1$. Using (\ref{eq:2-8.1}), equation (\ref{eq:2-8}) takes the form
\begin{equation}
    \dfrac{d^{2}y}{d r^{2}}+ \left[\dfrac{1}{r}+ \dfrac{1-2i \omega }{r-1}-2i \omega \right] \dfrac{dy}{dr}- \left[\dfrac{2i\omega + l(l+1)}{r-1}- \dfrac{l(l+1)}{r} \right]y_{\omega l}(r)=0.
\label{eq:2-9}
\end{equation}
Recall that the standard form of the Heun confluent equation (see equation (\ref{eq:a1})) is made up of 5 parameters, 2 regular singularities $(r=0\mbox{ and }r=1)$ and one irregular one ($r=\infty$). The equation (\ref{eq:2-9}) has precisely the standard form and shares the $3$ singularities. When comparing it with the equation (\ref{eq:a1}), we identify the $5$ parameters that characterize it as follows:
\begin{equation*}
    \gamma=1\mbox{, } \delta= 1-2i\omega \mbox{, } \beta=2i \omega \mbox{, } \alpha=1\mbox{ and }q=-l(l+1).
\end{equation*}

Two linearly independent solutions can be defined at the regular singularity $r = 1$, expressed using the confluent Heun function
\begin{equation}
    R^{I}(r,1)=r y^{I}_{\omega l}(r,1)=r \mbox{HeunC}\left[-q+\alpha \beta, \alpha \beta ,\delta, \gamma,\beta,1-r\right],
\label{eq:2-10}
\end{equation}
\begin{equation}\label{eq:2-11}
\begin{split}
    R^{II}(r,1)=& r y^{II}_{\omega l}(r,1)=r(r-1)^{1-\delta} \mbox{HeunC}\left[-q+\alpha \beta + (\beta -\gamma)(1-\delta),\right.\\
    &\left.\alpha \beta +\beta(1-\delta),2-\delta, \gamma,\beta,1-r\right].
\end{split}
\end{equation}

The two solutions are independent and show different behavior near the horizon. The first solution remains regular and maintains a constant value at that location.  Instead, the second solution makes an infinite number of turns on the unit circle in the complex plane and does not have a well-defined phase (see \cite{philipp2015analyticsolutionswaveequations}).

\section{Stochastic Trajectories with the Schwarzschild Metric}
\label{ch4}
In this section, we investigate the stochastic trajectories of quantum particles in the gravitational field of a Schwarzschild black hole within the framework of Stochastic Quantum Mechanics in curved spacetime. The previous section allowed us to determine the solution for a massless particle at the event horizon of a Schwarzschild black hole. Now, we can couple the ingoing (E-F) metric (\ref{eq2:5})
to the systematic and stochastic velocity (see equations (\ref{Pi sin e}) and (\ref{u relativista})) to solve the corresponding stochastic differential equations for the coordinates $x^{\mu}=(v, r, \theta, \phi)$ and thus obtain the particle trajectories. For massless particles, we propose $\mu_{0}=m=\omega_{0}$\footnote{Using units, we would have $\mu= m c/\hbar=\omega_{0}/c$.}, where $\omega_{0 }$ is the frequency of the particle. From equation (\ref{eq:madelungUnits}), we see that, in general, $S$ is defined by\footnote{For Eddington–Finkelstein coordinates we identify $x^{0}=v$
}
\begin{equation}
    S= \omega_{0} x^{0}- i  \mbox{ln}\left[\dfrac{\Phi}{\sqrt{\Phi \Phi^{*}}}\right].
\end{equation}
The field equation that describes a quantum particle immersed in a curved and fluctuating spacetime is precisely the Klein-Gordon equation. The solution $\Phi$ of this equation serves as the function that determines the stochastic trajectory due to the influence of the surrounding. Therefore, we now consider the solution (\ref{eq:2-3}) as the function used to define these stochastic trajectories. Thus, by taking the (E-F) ingoing coordinates, we obtain $S$, which appears as follows
\begin{equation}
S=  \left(\omega_{0}-  \omega \right) v- \dfrac{i}{2}\mbox{ln}\left(\dfrac{\mbox{HeunC}}{\mbox{HeunC}^{*}}\right)+m\phi,
\end{equation}
where we have abbreviated the Heun Confluent function to “$\mbox{HeunC}$” (the same for the complex conjugate “$\mbox{HeunC}^{*}$”) and $m$ is associated with the spherical harmonics, $m=-l ,-l+1,…,l-1,l$.

The velocity of an individual particle $(v^{\mu})$ is constructed from the first term of (\ref{Pi sin e}). We obtain
\begin{equation} \label{eq:2-12}
\begin{split}
v^{\mu}=& \left[-\dfrac{i }{2 \omega_{0}} \partial_{r}\mbox{ln}\left(\dfrac{\mbox{HeunC}}{\mbox{HeunC}^{*}}\right),\right.\\
& \left. \dfrac{\left( \omega_{0}-  \omega\right)}{\omega_{0}} -1-\dfrac{i }{2 \omega_{0}} \left(1- \dfrac{1}{r} \right) \partial_{r} \mbox{ln}\left(\dfrac{\mbox{HeunC}}{\mbox{HeunC}^{*}}\right),0, \dfrac{m }{r^{2} \mbox{sin}^{2}\theta \omega_{0}}\right].
\end{split}
\end{equation}
Given $v^{\mu}$, we construct $\pi^{\mu}$ with (\ref{Pi sin e}):
\begin{equation} \label{eq:2-13} 
\begin{split}
\pi^{\mu}=& \dfrac{1}{2 \omega_{0}}\left[-i\partial_{r}\mbox{ln}\left(\dfrac{\mbox{HeunC}}{\mbox{HeunC}^{*}}\right),\right.\\
& \left. -i\left(1- \dfrac{1}{r} \right) \partial_{r} \mbox{ln}\left(\dfrac{\mbox{HeunC}}{\mbox{HeunC}^{*}}\right)- 2\omega,0, \dfrac{2 m }{r^{2} \mbox{sin}^{2}\theta }\right].
\end{split}
\end{equation}
The stochastic velocity (see (\ref{u relativista})) is constructed with the particle density $n$; namely
\begin{equation}
    n= \Phi \Phi^{*}=\mbox{HeunC}\mbox{ HeunC}^{*}\left(Y_{l}^{m}(\theta,0)\right)^{2}.
\end{equation}
Performing the corresponding calculations, we found
\begin{equation} \label{eq:2-14}
\begin{split}
u^{\mu}=& \dfrac{1}{2 \omega_{0}}\left[ \partial_{r}\left(\mbox{HeunC} \mbox{ HeunC}^{*}\right),\right.\\
& \left. \left(1- \dfrac{1}{r} \right) \partial_{r}\left(\mbox{HeunC} \mbox{ HeunC}^{*}\right),\dfrac{1}{r^{2}} \partial_{\theta} \left( Y_{l}^{m} (\theta,0)\right)^{2},0 \right].
\end{split}
\end{equation}

To find the stochastic trajectories, it is necessary to solve the differential equation (\ref{eq:Forward SDE}) o (\ref{eq:Backward SDE}). We have chosen to use a forward stochastic differential equation, since we want to obtain solutions future evolutions, that lead to trajectories that goes into the event horizon. The Backward equations would only draw the stochastic trajectories that lead into the actual position of the particle, that is the possible different trajectories that lead into the 'present'. Then for this case the deterministic part $\mathcal{U}^{\mu}$, is defined as $\mathcal{U}^{\mu}= \pi^{\mu}+ u^{\mu}$ and its result is
\begin{equation} \label{eq:2-15}
\begin{split}
\mathcal{U}^{\mu}_{+}=& \dfrac{1}{2 \omega_{0}}\left[-i \partial_{r} \mbox{ln} \left(\dfrac{\mbox{HeunC}}{\mbox{HeunC}^{*}} \right)+\partial_{r}\left(\mbox{HeunC} \mbox{ HeunC}^{*}\right),\right.\\
& \left. \left(1- \dfrac{1}{r} \right) \left(i \partial_{r} \mbox{ln} \left(\dfrac{\mbox{HeunC}}{\mbox{HeunC}^{*}} \right)-\partial_{r}\left(\mbox{HeunC} \mbox{ HeunC}^{*}\right)\right)-2\omega,\dfrac{1}{r^{2}} \partial_{\theta} \left( Y_{l}^{m} (\theta,0)\right)^{2}, \dfrac{2 m}{r^{2} \mbox{sin}^{2}\theta} \right].
\end{split}
\end{equation}
The Heun Confluent function is a complex function, and to solve the respective stochastic differential equation we use Mathematica in its 14.1 edition. In order for the program to process the calculation of the HeunC function, it is necessary to divide it into its real and imaginary parts
\begin{equation}
    \mbox{HeunC}= \mbox{Re}[\mbox{HeunC}]+i\mbox{Im}[\mbox{HeunC}]=\mbox{R}[\mbox{H}]+i\mbox{I}[\mbox{H}],
\label{eq:2-16}
\end{equation}
\begin{equation}
    \mbox{HeunC}^{*}= \mbox{Re}[\mbox{HeunC}]-i\mbox{Im}[\mbox{HeunC}]=\mbox{R}[\mbox{H}]-i\mbox{I}[\mbox{H}].
\label{eq:2-17}
\end{equation}
Replacing equations (\ref{eq:2-16}) and (\ref{eq:2-17}) into equation (\ref{eq:2-15}), we arrive at
\begin{equation} \label{eq:2-18}
\begin{split}
\mathcal{U}^{\mu}_{+}=& \dfrac{1}{2 \omega_{0}}\left[-2 \left(\dfrac{\mbox{R}[\mbox{H}']\left(\mbox{I}[\mbox{H}]-\mbox{R}[\mbox{H}]\right)-\mbox{I}[\mbox{H}']\left(\mbox{I}[\mbox{H}']+\mbox{R}[\mbox{H}] \right)}{\mbox{R}^{2}[\mbox{H}]+\mbox{I}^{2}[\mbox{H}']} \right),\right.\\
& \left. -2 \left\{\left(1- \dfrac{1}{r} \right) \left(\dfrac{\mbox{R}[\mbox{H}']\left(\mbox{I}[\mbox{H}]-\mbox{R}[\mbox{H}]\right)-\mbox{I}[\mbox{H}']\left(\mbox{I}[\mbox{H}']+\mbox{R}[\mbox{H}] \right)}{\mbox{R}^{2}[\mbox{H}]+\mbox{I}^{2}[\mbox{H}']}\right)-\omega \right\},\dfrac{1}{r^{2}} \partial_{\theta} \left( Y_{l}^{m} (\theta,0)\right)^{2}, \dfrac{2 m}{r^{2} \mbox{sin}^{2}\theta} \right].
\end{split}
\end{equation}
The associated diffusion coefficient in the system is $\sigma=1/2m=1/2\omega_{0}$\footnote{If we use units, we would have $\sigma=\hbar/2m=c^{2}/2\omega_{0}$.}. As open source, the programming code can be consulted at \cite{JuanJerezRepo2024}. Each stochastic differential equation has an independent Wiener process (in particular, we will solve the coupled differential equations of $r(\tau)$ and $v(\tau)$). The stochastic process is defined using $\mbox{\textbf{ItoProcess}}$, which describes the evolution $r(\tau)$ and $v(\tau)$. $\mbox{\textbf{RandomFunction}}$ is used to generate stochastic trajectories of the Itô process on a time interval $\tau$ and with a time step of $0.01$. The two solutions are first plotted as functions of the parameter $\tau$. A parametric curve $r(\tau)\mbox{ Vs } v(\tau)$ is then constructed, and finally the results are presented again in the original coordinates $r \mbox{ Vs } t$.

We begin by varying the angular momentum quantum number $l$, taking the values $l=1, l=2$, and $l=3$. The initial condition for the coordinate $v$ is selected by setting $t=0$ and specifying the initial radial position (recall that the Schwarzschild radius is $r_{s}=1$). We adopt $\omega_{0}=\omega=35$, which physically corresponds to $\omega_{0}\approx 3\times 10^{18} \mbox{Hz}$, that is, a frequency in the X-ray range. The value of $\omega_{0}=35$ is chosen to emulate the value of $\sigma$ with conventional units. All simulations were performed in the interval $[0,1]$ of the parameter $\tau$.

The panel Fig. (\ref{fig:multipanel-trajectories_1}) contains all the simulation information. It is divided into three rows (from top to bottom). In the top row, each graph contains the trajectories $r(\tau)$ and $v(\tau)$ with respect to the parameter $\tau$, establishing 7 trajectories for each solution. The initial condition of $v(\tau)$ causes this trajectory to start at a different position than that of $r(\tau)$. It is also worth noting that if the radial initial condition were set closer to the black hole, the time trajectory would start at negative values (which is expected, since the shape of the transformation allows it). Regarding the behavior of the solutions, $v(\tau)$ with respect to $\tau$ exhibits significantly larger fluctuations than the radial trajectories, especially for $l=1$ and $l=2$. The radial dynamics show a robust pattern: all trajectories reach the event horizon, although they do so with different values of the parameter $\tau$. For $l=1$, some radial trajectories take less time to reach the event horizon compared to trajectories with $l=2$ and $l=3$. While this behavior is not observed identically in all numerical runs, it is an expected result: photons with lower angular momentum (as at $l=1$) experience less angular influence, whereas for higher values of $l$, the angular contribution introduces a more complex structure to the radial fall.

The parametric plots $r(\tau) \mbox{ Vs } v(\tau)$ for each $l$ are shown in the middle row of Fig. (\ref{fig:multipanel-trajectories_1}). These plots illustrate how the radial and time coordinates evolve simultaneously along each stochastic realization. All trajectories are directed toward the black hole singularity (none escape the black hole), and each is represented in a different color.

Eddington–Finkelstein coordinates allow for the unambiguous description of trajectories that cross the event horizon, something that is not possible in Schwarzschild coordinates due to the coordinate singularity at $r = 2M$. However, the parameter $v$—the advanced time—does not correspond to the time measured by a stationary observer at infinity; rather, it is better suited for describing incoming fields or trajectories near the horizon. For this reason, it is convenient to return to the original coordinates $(t,r,\theta,\phi)$ when interpreting the evolution from the perspective of a distant observer. This is described by the final row in Fig. (\ref{fig:multipanel-trajectories_1}). All the graphs show the same qualitative behavior: the coordinate singularity at the event horizon and the stochastic trajectories progressively approaching it. Trajectories drawn sufficiently close to the horizon "freeze" as $r\rightarrow 2M$, reflecting that a distant observer assigns an infinite coordinate time to crossing the horizon. Consequently, such trajectories do not cross the horizon but instead asymptotically approach it in infinite coordinate time, as predicted by classical Schwarzschild geometry. Although curves within the horizon are also shown, such extensions should not be interpreted physically in this coordinate system.

\begin{figure}[H]
    \centering
    % Ajusta el ancho si deseas más grande o más pequeño
    \includegraphics[width=1.04\textwidth]{Imagenes/MultiPanel_9Figs_Final.png}
    \caption{Multipanel visualization of the stochastic trajectories corresponding to the angular momentum values $l=1,2,3$. \textbf{Top row}: Radial trajectories $r(\tau)$ (solid red) and time trajectories $v(\tau)$ (dashed blue) as a function of the affine parameter $\tau$, along with the event horizon $r=1$ (black horizontal line) and the first horizon crossing marker (green dot). \textbf{Middle row}: Parametric plots $r(\tau)$ versus $v(\tau)$, showing the coupled stochastic evolution of both variables for each value of $l$. \textbf{Bottom row}: Radial coordinate $r$ as a function of time $t$, illustrating how all trajectories asymptotically approach the event horizon, depending on the angular momentum. In the middle and bottom rows, the trajectories are represented with different colors, and the event horizon $r=1$ (black horizontal line) is included again. Each graph contains seven realizations of the stochastic trajectories.
    }
    \label{fig:multipanel-trajectories_1}
\end{figure}

\begin{figure}[H]
    \centering
    % Ajusta el ancho si deseas más grande o más pequeño
    \includegraphics[width=1.04\textwidth]{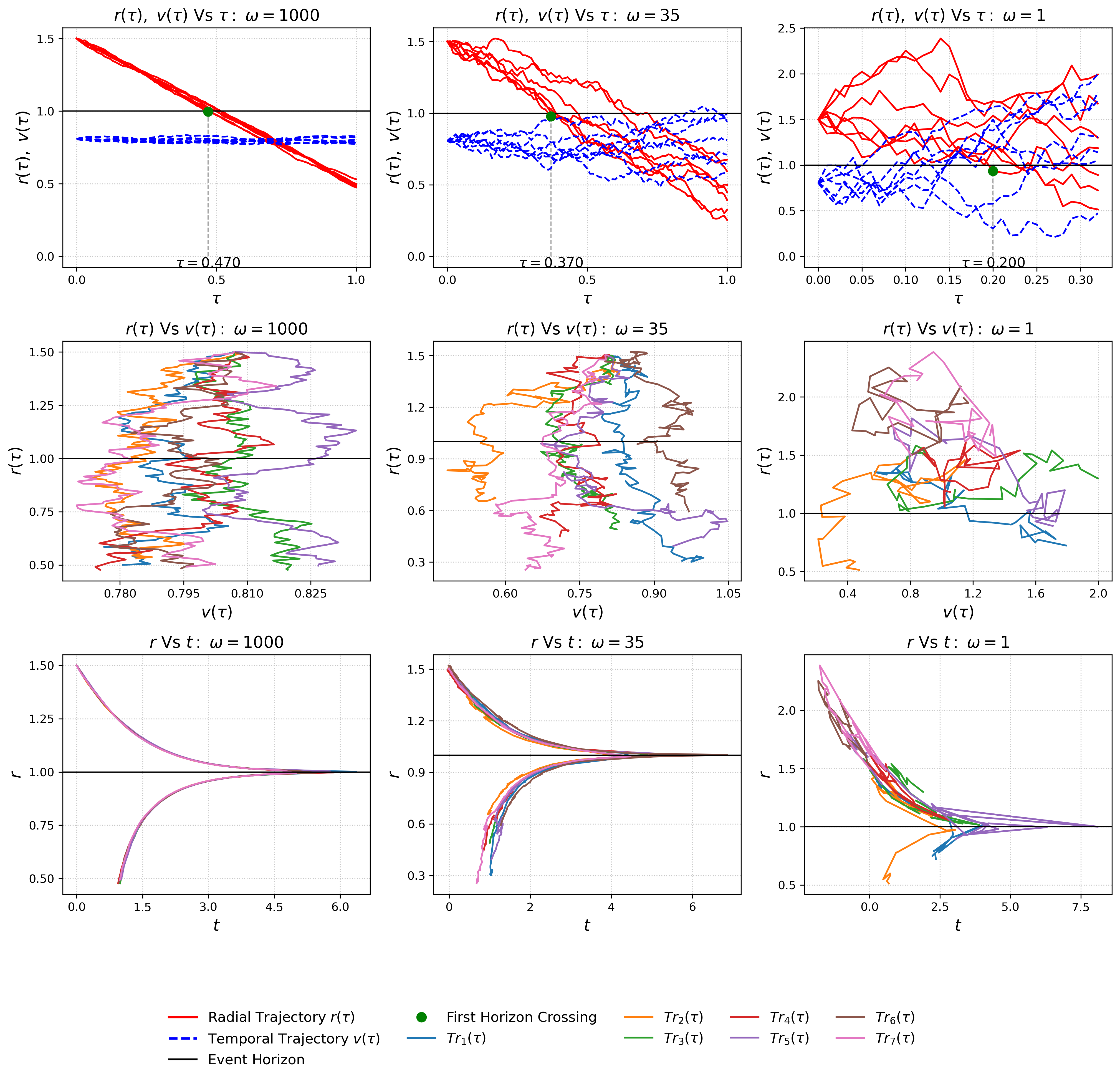}
    \caption{Multipanel visualization of the stochastic trajectories corresponding to the frequency values $\omega_{0}=1000, 35, 1$. \textbf{Top row}: Radial trajectories $r(\tau)$ (solid red) and time trajectories $v(\tau)$ (dashed blue) as a function of the affine parameter $\tau$, along with the event horizon $r=1$ (black horizontal line) and the first horizon crossing marker (green dot). \textbf{Middle row}: Parametric plots $r(\tau)$ versus $v(\tau)$, showing the coupled stochastic evolution of both variables for each value of $\omega_{0}$. \textbf{Bottom row}: Radial coordinate $r$ as a function of time $t$, illustrating how all trajectories asymptotically approach the event horizon, depending on the angular momentum. In the middle and bottom rows, the trajectories are represented with different colors, and the event horizon $r=1$ (black horizontal line) is included again. Each graph contains seven realizations of the stochastic trajectories.
    }
    \label{fig:multipanel-trajectories_2}
\end{figure}

The next parameter we will vary is $\omega_{0}$ (and, along the same lines, $\sigma$). The corresponding simulations are shown in the Fig.  (\ref{fig:multipanel-trajectories_2}), arranged in the same way as in the previous panel and using the same initial conditions. Physically, we use X-rays with a frequency of approximately $\omega_{0}\approx 3\times 10^{18} \mbox{Hz}$ (which in natural units, $\omega_{0}=35$). In our stochastic model, as the frequency increases, the diffusion coefficient decreases, so the trajectories become progressively more deterministic since the random contribution will be small. This can be seen in the top row for the case $\omega_{0}=1000$ (a value used only as an example, since physically it would represent a particle with too much energy), where the trajectories are smoother and their variability is noticeably lower. Conversely, as the frequency decreases, the stochastic contribution increases. The cases $\omega_{0}=35$ and $\omega_{0}=1$ (corresponding to less energetic X-rays) show, compared to the previous figures, greater fluctuations in the solutions. We do not consider even lower frequencies because, in our numerical scheme, greater diffusion implies a significant increase in computational complexity. Integrating the stochastic differential equations is particularly costly, since at each step the real and imaginary parts of the HeunC function must be evaluated. Consequently, the integration interval at $(\tau)$ is reduced (as observed for $\omega_{0}=1$).

The middle and bottom rows of the panel retain the same interpretative structure as in Fig. (\ref{fig:multipanel-trajectories_1}). where the parametric graphs and trajectories are found in the coordinates $(t,r)$ for different $\omega_{0}$, respectively.

\begin{figure}[H]
    \centering
    % Ajusta el ancho si deseas más grande o más pequeño
    \includegraphics[width=1.04\textwidth]{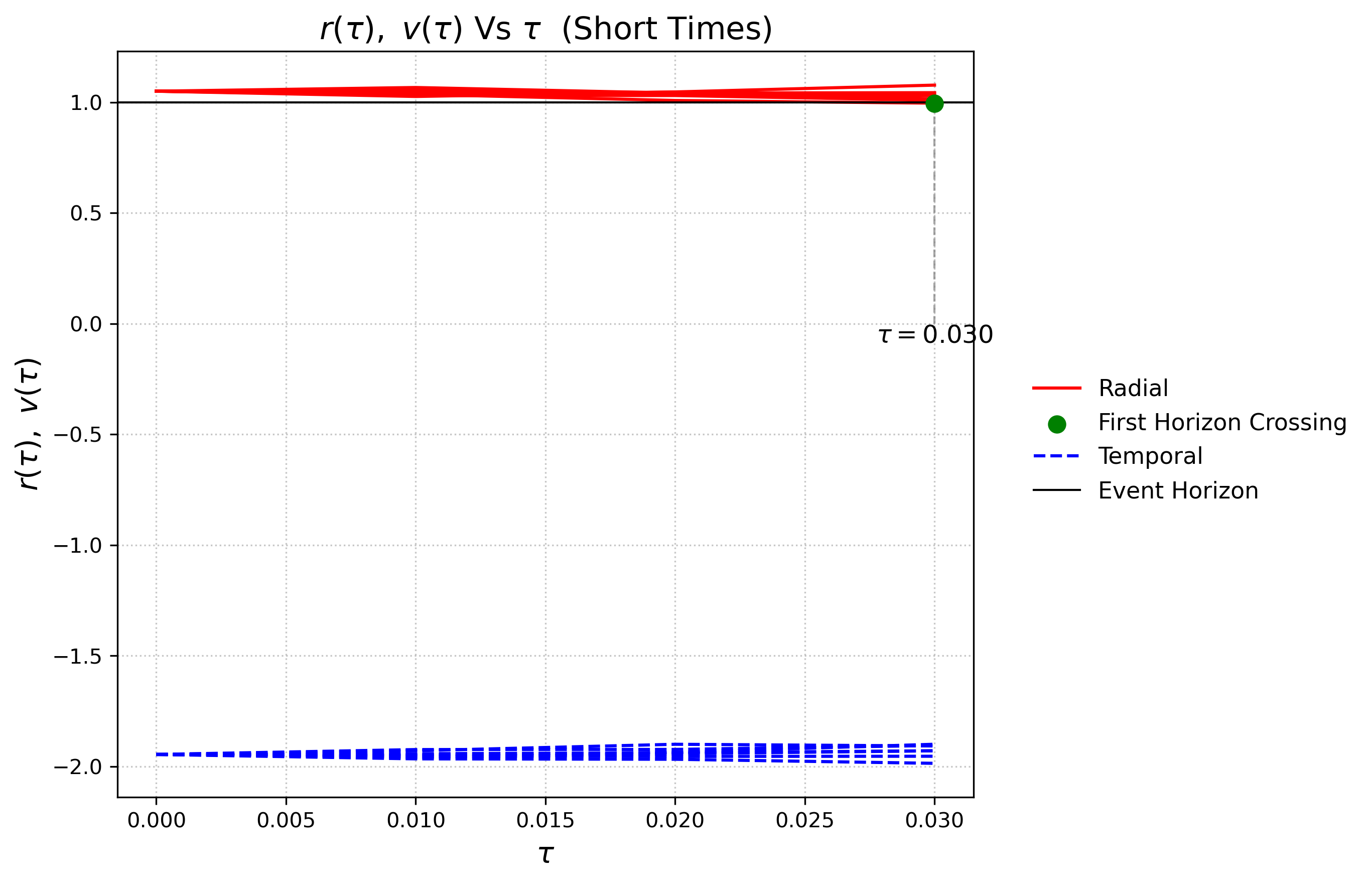}
    \caption{Visualization of coordinates, $r(\tau)$ and $v(\tau),$ evolution for short times, the corresponding parameters are $\omega_0=35$, the initial point is $r(0)=1.05r_s$. The event horizon is located at $r=1,$ the times goes from $0$ to $0.03$. 
    }
    \label{fig:multipanel-trajectories_3}
\end{figure}

The last figure, Fig. (\ref{fig:multipanel-trajectories_3}), is of great importance in the physical formalism. We show the evolution of the coordinates at short times, and we notice that the stochasticity is low. This is in accordance with the expected from stochastic formalism. For small times the stochasticity does not play an important role, the deterministic contribution domain the particle response over the random, because locally in spacetime we have independent stochastic processes, leading to not violating the Lorentz invariance as suggested in \cite{serva2021brownian}. This important feature is relevant in the stochastic formalism, as the stochastic differential equations are not invariant; but for short times we can build invariants, while for long times the stochastic dynamics rule over the deterministic leading to invariance violations. However, the obtained Klein-Gordon equation is invariant, in accordance of GR.

%Aquí conectas con lo anterior (es decir, acá escribes lo de tiempos coortos). Los parametros fueron: \[Omega]i = 35; r0 = 1.05 rs; oluEF = RandomFunction[EFITO, {0, 0.03, 0.01}, 7, Method -> "StochasticRungeKutta"]; 
\section{Conclusions}
\label{ch5}
In this work, an approach to stochastic quantum mechanics in curved spaces is developed, using the methodology presented by Chavanis in \cite{chavanis2016derivationgeneralizedschrodingerequation,axioms13090606} extended to the relativistic formalism followed by Escobar, Matos, and Aquino \cite{escobar2025fundamental}. The effects of spacetime curvature on quantum trajectories are analyzed in detail, applying this stochastic formalism to the particular case of the Schwarzschild black hole.

The analysis demonstrates that the stochastic formalism reproduces the Klein–Gordon equation and thus captures the full quantum dynamics of a scalar particle in curved spacetime. In this framework, stochastic fluctuations of the spacetime geometry influence the motion of particles, giving rise to stochastic trajectories that encode quantum behavior.

To explore these effects explicitly, we study configurations of massless particles in the vicinity of a Schwarzschild black hole using incoming Eddington-Finkelstein coordinates. The radial equation directly takes the form of the confluent Heun differential equation. This solution, represented as a Frobenius series expansion, is implemented in the program Wolfram Mathematica 14.1, which we will use as a reference for all simulations. The angular equation provides the spherical harmonics, and the temporal equation is solved using a classical exponential function. 

The general solution obtained allows us to calculate fundamental quantities within the stochastic formulation of quantum mechanics in curved spacetimes. This determines the hydrodynamic and stochastic velocities of the system and, consequently, the stochastic trajectories of particles near the event horizon. Under specific initial conditions for the photon (in this case, X-rays), the numerical results show that the stochastic formalism is entirely consistent with the relativistic structure of the problem. Itô's independent realizations for the variables $r(\tau)$ and $v(\tau)$ exhibit consistent behavior: (i) all radial trajectories reach the horizon for every value of angular momentum $l$ and frequency $\omega_{0}$, where low values of $l$ produce a faster approximation and high values of $\omega_{0}$ generate more deterministic trajectories (while lower frequencies increase randomness); (ii) upon returning to Schwarzschild coordinates, the trajectories show the asymptotic approximation to the horizon $r_{s}=2M$, consistent with the fact that a distant observer assigns an infinite coordinate time to cross it; and (iii) at short times, where stochastic fluctuations are negligible, the dynamics are dominated by the geodesic contribution.

The results suggest that stochastic quantum mechanics offers an alternative approach to exploring quantum effects in a gravitational context at particles' level. While this work focuses on the Schwarzschild metric, the methods and findings are applicable to various gravitational metrics and scenarios. This paves the way for future research in broader contexts, such as Kerr metrics, or environments with more complex gravitational perturbations.

\section*{Acknowledgements}
JSJR and ESEA gratefully acknowledge the financial support of the SECIHTI scholarships.
This work was partially supported by SECIHTI M\'exico under grants CBF-2025-G-1720 and CBF-2025-G-176. Also  by the grant I0101/131/07 C-234/07 of the Instituto Avanzado de Cosmolog\'ia (IAC) collaboration (http://www.iac.edu.mx/), and for the computing time granted by LANCAD and SECIHTI in the Supercomputer Hybrid Cluster "Xiuhcoatl" at GENERAL COORDINATION OF INFORMATION AND COMMUNICATIONS TECHNOLOGIES (CGSTIC) of CINVESTAV.
URL: http://clusterhibrido.cinvestav.mx/

\printbibliography

\appendix

\section{Confluent Heun Differential Equation}
\label{chap: Appendix A}

The confluent Heun differential equation (a linear, homogeneous, second-order ordinary differential equation), as presented in \cite{philipp2015analyticsolutionswaveequations}, has the following form:
\begin{equation}
    \left[ \dfrac{d^{2}}{dz^{2}}+ p_{1}(z) \dfrac{d}{dz} +p_{0}(z)\right] y(z)=0
\label{eq:a1}
\end{equation}
where $p_{i}(z)$ are functions of rational coefficients and $z$ is a point on the extended complex plane, also known as the Riemann sphere, which includes $z = \infty$.

When all singular points of a differential equation are regular, it is classified as a Fuchsian equation (otherwise, it is known as a confluent equation). The confluent Heun equation (CHE) belongs to the larger family of Heun equations and arises when two regular singularities of the general Heun equation (GHE) merge, forming an irregular singularity at $z = \infty$. Its standard form (see (\ref{eq:a1})), is accompanied by:
\begin{equation}
    p_{1}(z)= \dfrac{\gamma}{z}+\dfrac{\delta}{z-1}-\beta,
\end{equation}
\begin{equation}
    p_{0}(z)= \left(\dfrac{\alpha \beta-q}{z-1}+ \dfrac{q}{z} \right).
\end{equation}
The CHE has regular singularities at $z = 0,1$ and an irregular singularity at $z = \infty$. Around the regular singularities we can build local Frobenius-type solutions. For the irregular singularity, we can build Thome-type solutions. For the singularity $z=0$ we find
\begin{equation}
    y^{I}(z;0)=\sum_{k=0}^{\infty}a_{k}z^{k},
\end{equation}
\begin{equation}
    y^{II}(z;0)=\sum_{k=0}^{\infty}b_{k}z^{k+1-\gamma}.
\end{equation}
For $z=1$, the solution is
\begin{equation}
        y^{I}(z;1)=\sum_{k=0}^{\infty}c_{k}(z-1)^{k},
\end{equation}
\begin{equation}
        y^{II}(z;1)=\sum_{k=0}^{\infty}d_{k}(z-1)^{k+1-\delta}.
\end{equation}
The coefficients of each solution can be found by replacing the solutions in the differential equation (\ref{eq:a1}). We will find two linearly independent solutions, which we call $I$ and $II$. To know in which singularity we are located, we take the notation $y^{I,II}(z;\cdot)$, with $(\cdot)$ indicating the singularity. These solutions are convergent within a circle in a complex plane, centered at the respective regular singularity with a radius that is the distance to the next neighboring singular point \cite{philipp2015analyticsolutionswaveequations}. This standard form of the confluent Heun differential equation is not exactly the form that the Mathematica program handles as its standard form. However, the program itself rearranges its solution in such a way that, for the regular singularity $z=0$, , the standard confluent Heun function, with domain $|z|<1$, is
\begin{equation}
    y^{I}(z,0)=\mbox{HeunC}[-1,-\alpha\beta,\gamma,\delta,-\beta,z ],
\end{equation}
\begin{equation}
    y^{II}(z,0)=z^{1-\gamma}\mbox{HeunC}[-1+ (1-\gamma)(-\beta-\delta),-\alpha\beta-\beta(1-\gamma),2-\gamma,\delta,-\beta,z ].
\end{equation}
and for the singularity $z=1$, is
\begin{equation}
y^{I}(z,1)=\mbox{HeunC}[-q+\alpha \beta,\alpha\beta,\delta,\gamma,\beta,1-z],   
\end{equation}
\begin{equation}
y^{II}(z,1)=(z-1)^{1-\delta}\mbox{HeunC}[-q+\alpha \beta+(\beta-\gamma)(1-\delta),\alpha\beta+\beta(1-\delta),2-\delta,\gamma,\beta,1-z].
\end{equation}
Solutions to the irregular singularity $(z=\infty)$ are created using the Thóme solution. Such solutions look like this:
\begin{equation}
    y^{I}(z;\infty)=\sum_{k=0}^{\infty}\rho_{k} z^{-(k+\alpha)}
\end{equation}
\begin{equation}
    y^{I}(z;\infty)=e^{\beta z}\sum_{k=0}^{\infty}\sigma_{k} z^{-(k+\gamma+\delta-\alpha)}.
\end{equation}
Again, the coefficients can be found by introducing the solutions of the differential equation (\ref{eq:a1}). A way of relating all the solutions is given in \cite{philipp2015analyticsolutionswaveequations}.

\end{document}